\documentclass[aps,prresearch,twocolumn]{revtex4-2} 

\usepackage[]{graphicx}
\usepackage[dvipsnames]{xcolor}
\usepackage[colorlinks=true,citecolor=NavyBlue,linkcolor=RubineRed,urlcolor=Cerulean,pdfencoding=auto]{hyperref}
\usepackage[]{amsmath}
\usepackage[]{amssymb}

\def\bra#1{\mathinner{\langle{#1}|}}
\def\ket#1{\mathinner{|{#1}\rangle}}
\def\braket#1{\mathinner{\langle{#1}\rangle}}

\def\abs#1{\mathinner{|{#1}|}}

\def\proj#1{\ket{#1}\bra{#1}}
\def\op#1#2{\ket{#1}\bra{#2}}

\def\EM{E_{\mathrm{max}}}
\def\G#1{\int_0^{#1} d(\tau) d\tau}

\begin{document}
\title{Efficient High-Fidelity Flying Qubit Shaping}

\author{Benedikt Tissot}
\email[]{benedikt.tissot@uni-konstanz.de}
\affiliation{Department of Physics, University of Konstanz, D-78457 Konstanz, Germany}

\author{Guido Burkard}
\email[]{guido.burkard@uni-konstanz.de}
\affiliation{Department of Physics, University of Konstanz, D-78457 Konstanz, Germany}

\begin{abstract}
  Matter qubit to traveling photonic qubit conversion is the cornerstone of numerous quantum technologies such as distributed quantum computing, as well as several quantum internet and networking protocols.
  We formulate a theory for stimulated Raman emission which is applicable to a wide range of physical systems including quantum dots, solid state defects, and trapped ions, as well as various parameter regimes.
  We find the upper bound for the photonic pulse emission efficiency of arbitrary matter qubit states for imperfect emitters and show a path forward to optimizing the fidelity.
  {Based on these results we propose a paradigm shift from optimizing the drive to directly optimizing the temporal mode of the flying qubit using a closed-form expression.}
  Protocols for the production of time-bin encoding and spin-photon entanglement are proposed.
  Furthermore, the mathematical idea to use input-output theory for pulses to absorb the dominant emission process into the coherent dynamics, followed by a non-Hermitian Schr\"odinger equation approach has great potential for studying other physical systems.
\end{abstract}

\maketitle

\section{Introduction}
Efficient, tunable, and coherent quantum emitters are at the heart of many quantum technologies.
Prominent examples include entanglement distribution \cite{cirac97,leent22} as well as more general applications
for quantum networks and communication \cite{gisin07,nemoto14,munro15,grasselli19,zhang22,nadlinger22}
which can potentially enable a quantum internet \cite{kimble08,wehner18} with quantum-mechanically enhanced security and privacy.
Additionally, single-photon emission represents a cornerstone for several photonic technologies \cite{obrien09,giovannetti11,aspuru12,slussarenko19,brod19}.
Generally, there is a large interest in coherent quantum media conversion,
as it allows the connection between different quantum systems with diverse properties.
This enables hybrid quantum systems that combine the advantages of each subsystem.
Such hybrid quantum systems can combine matter systems with beneficial properties for storage or computation,
e.g. trapped ions \cite{bruzewicz19}, semiconductor qubits \cite{burkard23} implemented via quantum dots and defects in solids, or superconducting circuits \cite{kjaergaard20},
with easily transmittable photons \cite{slussarenko19}.
In this regard,
photons are the natural choice for traveling qubits \cite{bennett14} and
can be used to exchange quantum states or create entanglement between distant matter systems.

Cavity-enhanced stimulated Raman emission is an established technique
for controlled and (nearly) deterministic pulse emission, i.e., ``push button-like'' shaped pulse generation.
The on-demand emission promises a leap towards independence of emission and absorption
which is of the utmost importance when exchanging states between diverse systems.
The ability of Purcell enhancement to achieve a controllable emitter with high efficiency was already shown over a wide range of materials, e.g.
trapped ions as well as atoms \cite{kuhn02,duan03,morin19}, ``un-trapped'' atoms \cite{nisbet-jones11}, quantum dots \cite{sweeney14,pursley18}, and defects in solids \cite{sun18,knall22}.
Previous theoretical work focused on perfect emitters
\cite{cirac97,law97,gorshkov07,dilley12,vasilev10,khanbekyan17}.
%
\begin{figure}[!ht]
\centering
\includegraphics[width=\linewidth]{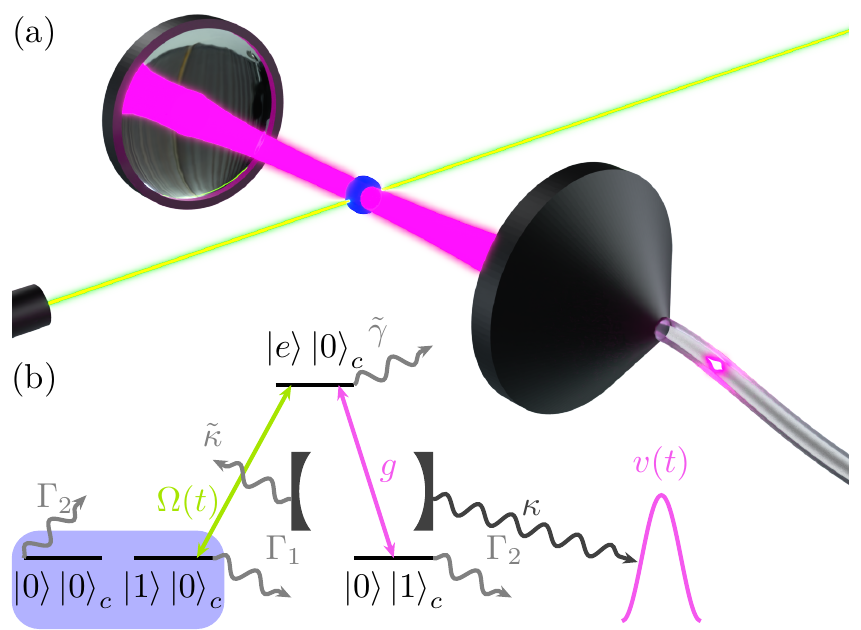} 
\caption{\label{fig:level}
  Illustration of the physical system (a) and energy level diagram (b) of a stimulated Raman emitter.
  The illustration depicts a quantum system (blue ball) coupled to an electromagnetic cavity mode (pink) as well as an excitation (e.g. laser) field (green).
  One of the mirrors couples out into a fiber enabling the emission of a photon pulse (pink glowing droplet).
  The level diagram (b) in the rotating frame depicts
  the excited state \(\ket{e}\) (ES) split from the two ground states (GS) \(\ket{0}, \ket{1}\) by the detuning $\Delta$ for the relevant states of the cavity $|0\rangle_c$ and $|1\rangle_c$ with zero and one photon.
  The lambda ($\Lambda$) system is set up by a controllable time-dependent (excitation field) Rabi amplitude $\Omega(t)$
  coupling $\ket{1}\ket{0}_c$ to the ES,
  as well as cavity interaction between the ES and $\ket{0}\ket{1}_c$ with single photon coupling strength \(g\).
  The cavity emits the photon wave packet with pulse shape $v(t)$ via the (right) out-coupling \(\kappa\), thus converting the matter qubit (shaded blue) to a flying qubit.
  Imperfections of the three-level system and cavity can lead to decoherence processes taken into account via the combined rates \(\tilde{\gamma}\), \(\Gamma_1\), \(\Gamma_2\), and $\tilde{\kappa}$.
}
\end{figure}

In this paper
we theoretically determine the fundamental fidelity bound of coherent state transfer for arbitrary pulse shapes from a stationary matter three-level system (3LS)
via a cavity to a traveling qubit pulse
which facilitates a distinct approach to maximize the state transfer fidelity.
In particular, we are interested in the transfer of a superposition of \emph{qubit states} \(\alpha_0 \ket{1} + \beta_0 \ket{0}\) 
via the excited state \(\ket{e}\) and cavity to the traveling photon (qubit) \(\alpha_0 \ket{1}_v + \beta_0 \ket{0}_v\), see Fig.~\ref{fig:level}.

Previously derived photon retrieval bounds \cite{gorshkov07,dilley12,muecke13,morin19}
depending only on the cavity decay rate and cooperativity of the emitter-cavity coupling
can greatly overestimate the bound we calculate
which
includes additional system features,
most prominently different decoherence processes of the emitter
as well as the \emph{temporal shape of the flying qubit} $v(t)$
and initial superposition states, fundamentally necessary to understand spin-photon entanglement.
Because the bound depends on the shape of the photon, it is suited to find optimized flying qubit shapes,
providing a {paradigm shift} from {approaches that aim to find the optimal drive, e.g.,} the shortcut to adiabaticity \cite{baksic17} approach as well as theories eliminating the propagating pulse completely \cite{cirac97}.
%

\begin{figure*}[ht!]
\centering
\includegraphics[width=.8\linewidth]{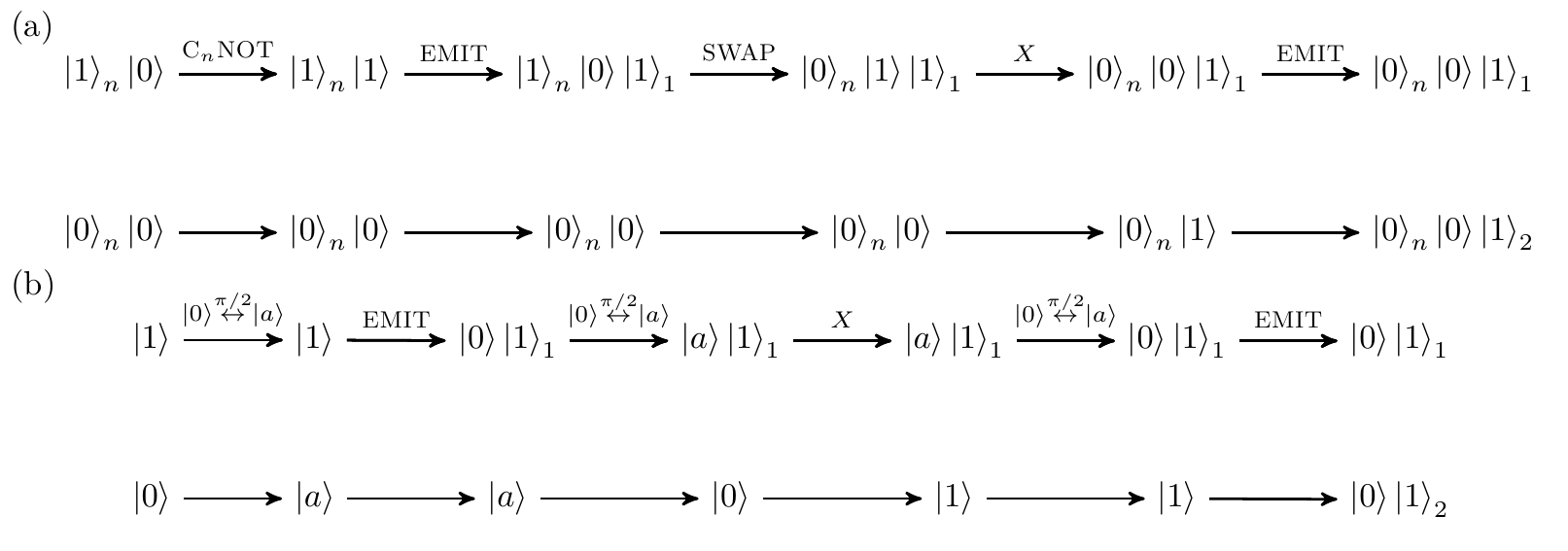} 
\caption[Emission of a time-bin qubit using an ancillary qubit or an ancillary state.]{\label{fig:timebin}
  Emission of a time-bin qubit using (a) an ancillary (e.g. nuclear spin) qubit or (b) an ancillary state.
  The main focus of this paper is the particular implementation of the \textsc{EMIT} process via cavity-enahanced stimulated Raman emission.
  $\mathrm{C}_{n}\mathrm{NOT}$ and \textsc{SWAP} refer to the two qubit gates between the matter qubit and the ancillary qubit and $X$ denotes the one-qubit $X$-gate of the matter qubit (or a \(\pi/2\)-pulse between \(\ket{0}\) and \(\ket{1}\)).
  Additionally, if one only implements the protocol up to the first emission, matter-photon entanglement is achieved.
}
\end{figure*}
The remainder of this paper is organized as follows.
First, we show how stimulated Raman emission can be used to generate spin-photon entanglement in Sec.~\ref{sec:stiment}.
In the following Sec.~\ref{sec:stimmodel} we introduce the model describing the emitter and the quantum pulse.
We present a closed-form solution of the dynamics in Sec.~\ref{sec:stimresults}
and introduce the temporal mode matching (Sec.~\ref{sec:stimtmm}) to link the emitter dynamics to the temporal mode.
We use the solution to bound the state transfer fidelity in Sec.~\ref{sec:stimbound} and show how to optimize the pulse shape to increase the fidelity in Sec.~\ref{sec:stimopt}.
We conclude the paper in Sec.~\ref{sec:stimconc}.

\section{Entanglement Generation via Stimulated Raman Emission}\label{sec:stiment}
Prior to the detailed analysis we point out quantum technological applications
of stimulated Raman emission beyond single photon sources,
e.g.,
to create a photon entangled with the matter qubit or to transfer the matter qubit to a time-bin qubit
if an additional long-lived matter state or (nuclear) spin ($n$) 
is available. 
For concreteness, we focus on the silicon-vacancy center in diamond, an established Raman emitter~\cite{sun18,knall22}
which features the silicon nuclear spin as a quantum memory.
Recently, a \(\textsc{C}_n\textsc{NOT}\) gate between the electronic qubit and the nuclear spin was demonstrated using microwaves (fidelity $\sim 99.9\%$)~\cite{stas22}.
After initializing the nuclear spin $\alpha_0 \ket{1}_n + \beta_0 \ket{0}_n$ state, we propose to use a nuclear \(\textsc{C}_n\textsc{NOT}\) followed by a Raman emission resulting in the entangled state $\alpha_0 \ket{1}_n \ket{1}_v + \beta_0 \ket{0}_n \ket{0}_v$.
The basic idea is to store the wave function amplitude \(\beta_0\) (of the qubit state \(\ket{0}\)) in an ancillary state during the emission from the qubit state \(\ket{1}\) to the first time bin \(\ket{1}_1\).

After the first emission, gates are applied between the qubit and the ancillary state(s) so that after another emission the qubit state is encoded in two time-bins, i.e., a time-bin qubit with the state \(\alpha_0 \ket{1}_1 + \beta_0 \ket{1}_2\).
The indices mark the time-bins which are independent pulses, e.g., pulses with non-overlapping envelopes.
In Fig.~\ref{fig:timebin} we show two examples of the implementation of this idea,
one for an ancillary qubit
(as is the case for the silicon nuclear spin in the {SiV}) 
and one for a single ancillary state.
For ancillary qubit implementation we need the two-qubit gates $\textsc{C}_n\textsc{NOT}$ and \textsc{SWAP}
while for the single extra state $\ket{a}$
we need the ability to apply a $\pi/2$ rotation between $\ket{0}$ and $\ket{a}$,
in both cases an $X$ gate between $\ket{0}$ and $\ket{1}$ is also required.
The \textsc{EMIT} gate corresponds to the stimulated Raman emission process.
In both cases, stopping the protocol after the first emission or replacing the \textsc{SWAP} by another $\textsc{C}_n\textsc{NOT}$ can be used to generate entanglement.
For the single ancillary state the time-bin entanglement can be achieved by omitting the second $\pi/2$ pulse for the single ancillary state.

The entangled state after the first emission is \( \alpha_0 \ket{1}_n \ket{1}_1 + \beta_0 \ket{0}_n \ket{0}_1\) using the ancillary qubit with states \(\ket{\sigma}_n\) (\(\sigma=0,1\))
or \(\alpha_0 \ket{0} \ket{1}_1 + \beta_0 \ket{a} \ket{0}_1\) using only one ancillary state \(\ket{a}\),
where \(\ket{m}_1\) with \(m=0,1,\dots\) is the number state of the first emitted pulse.
For the time-bin entanglement we find
\(\alpha_0 \ket{1}_n \ket{1}_1 + \beta_0 \ket{0}_n \ket{1}_2\)
and
\(\alpha_0 \ket{a} \ket{1}_1 + \beta_0 \ket{0} \ket{1}_2\).
Analogously, initializing the system in one of the ancillary states and then repeating a
(partial) transfer of the occupation to \(\ket{1}\) followed by an emission
enables time-bin qudit generation.

Applications for the protocol generating entanglement between a quantum memory and a flying qubit include entanglement exchange between distant matter nodes (heralded~\cite{barrett05} or combined with perfect absorption~\cite{dilley12,giannelli18} to ``pitch-and-catch'' \cite{cirac97})
and the generation of photonic cluster states~\cite{lindner09}.
Note that the following analysis is compatible with the outlined entanglement generation protocols because the analytical solution only makes assumptions about the initial preparation of the Raman emitter.

\section{Model}\label{sec:stimmodel}

In this scenario the quantum memory is better shielded
from decoherence than the emitter 
making it vital to understand and optimize the emission. 
The rotating frame Hamiltonian (see App.~\ref{app:LF} for additional details on the rotating frame)
of the cavity interacting with the {3LS} is
\begin{align}
 \label{eq:Hsys}
  H_S / \hbar
 =  \Delta \op{e}{e} + \left[ \Omega(t) \op{e}{1} + g c^{\dag} \op{0}{e} + \mathrm{H.c.} \right],
\end{align}
with the time-dependent Rabi frequency of the drive \(\Omega(t)\), the single photon coupling strength to the cavity $g$,
the detuning between the cavity and the $\ket{0}\leftrightarrow\ket{e}$ transition \(\Delta\),
the 3LS states \(\ket{0}, \ket{1}, \ket{e}\),
and the cavity photon annihilation operator \(c\).
The energy level structure is sketched in Fig.~\ref{fig:level}.

To include the emission from the cavity into a specific output pulse in the Hermitian dynamics
we employ the recently developed input-output theory for quantum pulses \cite{kiilerich19,kiilerich20}
using an open quantum systems approach.
Therefore, we can use
dissipators $L_i$ to model various incoherent processes
and 
rely on the Born-Markov approximation.
Since the relevant emission is into a pulse that travels away from the emitter,
this approximation is justified provided the cavity coupling \(\kappa\) is approximately constant for the spectrum of the pulse
and the emitted pulse varies slowly compared to the spectral range of the continuum field~\cite{kiilerich20}.
For a pulse with a carrier frequency in the optical range and durations in the order of \(0.1\,\)ns
or a microwave pulse and durations down to \(10\,\)ns,
we deem the assumptions well justified.

We take the input to be in the vacuum state and
employ input-output theory for pulses \cite{kiilerich19,kiilerich20} to model the emission (caused by directional single photon losses $c$ with rate $\kappa$) to the specific pulse as a \emph{virtual cavity} with time-dependent coupling to the system. In this way, the cavity would completely absorb the specific pulse.
This is achieved by a total Hamiltonian
\(H = H_S + \hbar \frac{i}{2} \sqrt{\kappa} \left[ g_v^{*}(t) c^{\dag} a - \mathrm{H.c.} \right]\),
with the cavity decay rate \(\kappa\), the annihilation operator of the virtual cavity \(a\) and time-dependent coupling strength \(g_v(t) = -{v^*(t)}/{\sqrt{\int_0^t dt'\, |v(t')|^2}}\) which is directly linked to the normalized pulse form \(v(t)\),
i.e. \(\int_0^T \abs{v(\tau)}^2 d\tau = 1\) with the pulse duration \(T\).
Combined with the dissipator
\(L_0(t) = g_v(t)^{*} a + \sqrt{\kappa} c\),
the total Hamiltonian leads to a cascaded evolution, resulting in a transfer of the quantum amplitudes from the emitter to the virtual cavity.
Additionally, a pulse shape that does not (perfectly) capture the dynamics of the emission process leads to incoherent losses via \(L_0(t)\);
in other words, only the emission into modes other than the temporal mode $v(t)$ are treated as losses.

To study the coherent state transfer of a matter state into the propagating wave packet,
we focus on the single excitation subspace using a non-Hermitian Hamiltonian approach
\cite{dalibard92,moelmer93,carmichael93,daley14},
described by the time-dependent Schr\"odinger equation
$i \hbar \frac{\partial}{\partial t} \ket{\Psi} = H_{\mathrm{NH}} \ket{\Psi}$
with \(H_{\mathrm{NH}} = H - \hbar \frac{i}{2} \sum_i L_i^{\dag} L_i\)
and using the ansatz wave function
\begin{align}
  \ket{\Psi(t)} =\, & \left[ \alpha(t) \ket{1} + \beta(t) \ket{0} + i \zeta(t) \ket{e} \right] \ket{0}_c \ket{0}_v \notag \\
  \label{eq:Psi}
  & + \eta(t) \ket{0} \ket{1}_c \ket{0}_v + \lambda(t) \ket{0} \ket{0}_c \ket{1}_v .
\end{align}
%
\section{Results}\label{sec:stimresults}
In the following we summarize the solution of the dynamics of the state transfer from the normalized state
$\ket{\Psi(0)} = (\alpha_0 \ket{1} + \beta_0 \ket{0}) \ket{0}_c \ket{0}_v$
to a state close to $\ket{\Psi_{\mathrm{target}}} = \ket{0} \ket{0}_c (\alpha_0 \ket{1}_v + \beta_0 \ket{0}_v)$.
Our first objective is to find a (closed-form) expression for the (approximate) fidelity of the state transfer
\begin{align}
  \label{eq:fidelity}
  F = \abs{ \braket{\Psi(T) | \Psi_{\mathrm{target}}} }^{2} = \abs{ \alpha_0^{*} \lambda(T) + \beta_0^{*} \beta(T) }^2,
\end{align}
as a function of the system parameters and the pulse shape.
One of the distinguishing features of our method is that it does not require the repeated numerical solution of any differential equations.

\subsection{Temporal Mode Matching}\label{sec:stimtmm}

First, we solve the amplitude that is not partaking in the emission and therefore is only subject to decoherence,
$\beta(t) = \beta_0 e^{-\Gamma_2 t /2}$.
Then we determine the optimal {relationship between the wave function amplitudes [Eq.~\eqref{eq:Psi}] and the pulse shape} by solving 
\(L_0(t) \ket{\Psi(t)} = 0\)
for the time-dependent coupling $g_v(t)$ to the virtual cavity.
Fulfillment of this condition implies that
the successful coherent emission is a no-jump trajectory in the model,
and yields
\begin{align}
  \label{eq:impedance-matching}
  g_v(t) = - \sqrt{\kappa} {\eta^{*}{(t)}}/{\lambda^{*}{(t)}} ,
\end{align}
which corresponds to the \emph{temporal mode matching} condition.
While the ideal process is fully coherent in our model,
the remaining incoherent processes ($L_i$, $i \ne 0$) describe unwanted errors.
%
Thus, we have immediate access to the probability of these errors
\(p_e = 1 - \braket{\Psi(t)|\Psi(t)} \ge 0\)
via the loss of the wave function norm,
underlining that the chosen approach is perfectly suited to describe the coherent transfer of population.
Within $p_e$ we here consider the combined decoherence rates
$\tilde{\gamma}, \Gamma_1, \Gamma_2$ respectively for the 3LS states  $\ket{e}, \ket{1}, \ket{0}$, as well as additional losses of the cavity $\tilde{\kappa}$.
In App.~\ref{app:decrate} we relate the combined decoherence rates to the corresponding dissipators of the Lindblad master equation modeling incoherent transitions, e.g., decay and dephasing processes.
We are focusing on efficient quantum emitters where
high fidelities are possible.
By employing the Cauchy-Schwartz inequality we find $p_e \le 1 - F$ such that
$p_e$ is small (\(p_e \ll 1\)) for suitable pulses
that reach high fidelities. 
This makes our approach highly accurate because the unwanted emission of multiple excitations
and the back action on the coherent trajectory after a quantum jump that are neglected within the non-Hermitian Schrödinger equation are highly unlikely.
This confirms the usefulness of the
non-normalized coherent
trajectory to calculate fidelities approaching unity.

The pulse shape can be linked to the dynamics via the temporal mode matching \eqref{eq:impedance-matching}.
Combined with the non-Hermitian Schrödinger equation, we arrive at a set of differential equations for the amplitudes, whose solution we calculate in App.~\ref{app:solvingHNH}.
We find
\begin{align}
  \label{eq:eta}
  \eta(t) = \, & E \alpha_0 e^{-\Gamma_2 t / 2} v(t) / \sqrt{\kappa} , \\
  \label{eq:lambda}
   \lambda(t) = \, & E \alpha_0 e^{- \Gamma_2 t / 2} \sqrt{ \int_{0}^{t} \abs{v(\tau)}^2 \, d\tau } ,
\end{align}
where the positive proportionality constant $E$ takes into account that the pulse shape is normalized independently of the dynamics
and 
we term it 
\emph{matter-photon (amplitude) conversion efficiency}
since it can be defined as the (renormalized) amplitude transfer ratio
$E= (\lambda(T) / \alpha_0) e^{ \Gamma_2 T / 2}$.
In turn, the fidelity~\eqref{eq:fidelity}
is directly bound by $\alpha_0^{*} \lambda(T)$
and thus by the maximal achievable $E_{\mathrm{max}}$.
To calculate $E_{\mathrm{max}}$, we apply the idea of photon shaping \cite{vasilev10},
where we solve the non-Hermitian dynamics in reverse by imposing $\eta, \lambda$, resulting in
\begin{align}
  \label{eq:zeta}
  \zeta(t) = \, & \frac{\Gamma_{2} + \kappa + \tilde{\kappa}}{2 g} \eta(t) + \frac{1}{g} \dot{\eta}(t) , \\
  \label{eq:Omega}
  \Omega(t) = \, & - \frac{( \tilde{\gamma}/2  + i \Delta ) \zeta(t) + g \eta(t) + \dot{\zeta}(t)}{\alpha(t)} .
\end{align}
The remaining amplitude evolves according to
\begin{align}
  \label{eq:alpha}
  \alpha(t) = \alpha_0  e^{i \phi(t) - \Gamma_1 t/2} \sqrt{1 - E^2 \G{t}} ,
\end{align}
with the phase evolution $\phi(t)$ and the (re-normalized) depletion rate $d(t)$
which are closed-form analytic functions of the pulse shape $v(t)$ and system parameters $g, \kappa, \tilde{\gamma}, \tilde{\kappa}, \Gamma_1, \Gamma_2$.
The depletion rate is
\begin{widetext}
\begin{align}
  d(t) = &\, e^{t \left(\Gamma_{1} - \Gamma_{2}\right)} \Bigg\{
  \left[ 1 +
 \frac{\tilde{\kappa}}{\kappa} + \frac{\tilde{\gamma} - \Gamma_2}{g^{2}} \left( \frac{\kappa}{4} \left( 1 + \frac{\tilde{\kappa}}{\kappa} \right)^2 + \frac{\dot{\theta}^2(t)}{\kappa} \right) + \frac{2 \ddot{\theta}(t) \dot{\theta}(t)}{\kappa g^2} \right] f^{2}(t)
    +  \frac{1}{g^{2}} \left( 1 + \frac{\tilde{\kappa}}{\kappa} \right) f(t) \ddot{f}(t) + \frac{2}{\kappa g^{2}} \dot{f}(t) \ddot{f}(t) \notag \\
  & + \left[ \frac{2}{\kappa} + \frac{\kappa}{2 g^{2}} \left( 1 + \frac{\tilde{\kappa}}{\kappa} \right)^2 + \frac{\tilde{\gamma}-\Gamma_2}{g^{2}} \left( 1 + \frac{\tilde{\kappa}}{\kappa} \right) + \frac{2 \dot{\theta}^2(t)}{\kappa g^2} \right] f(t) \dot{f}(t)
    + \left[ \frac{1}{g^{2}} \left( 1 + \frac{\tilde{\kappa}}{\kappa} \right) + \frac{\tilde{\gamma} - \Gamma_2}{\kappa g^{2}} \right] \dot{f}^2(t)
  \label{eq:dG}
              \Bigg\} ,
\end{align}
\end{widetext}
where we introduce the photon envelope phase \(\theta(t) \in \mathbb{R}\) and amplitude \(f(t) \in \mathbb{R}\), i.e., \(v(t) = e^{i \theta(t)} f(t)\).
The phase evolution \(\phi(t)\) can be found in App.~\ref{app:solvingHNH}.
We stress that $d(t)$ is independent of the detuning $\Delta$ and matter-photon conversion efficiency $E$.
On the other hand, 
the phase evolution $\phi(t)$ is an integral expression additionally depending on $\Delta$.
For a pulse with constant complex phase and for $\Delta = 0$
we have $\phi(t) = 0$.
We highlight that \(\Omega(t)\), {\(d(t)\),} and \(\phi(t)\) are all independent of the initial state of the matter qubit {($\alpha_0, \beta_0$)}.

\subsection{State Transfer Fidelity Bound}\label{sec:stimbound}
\begin{figure}
\centering
\includegraphics[width=8.5cm]{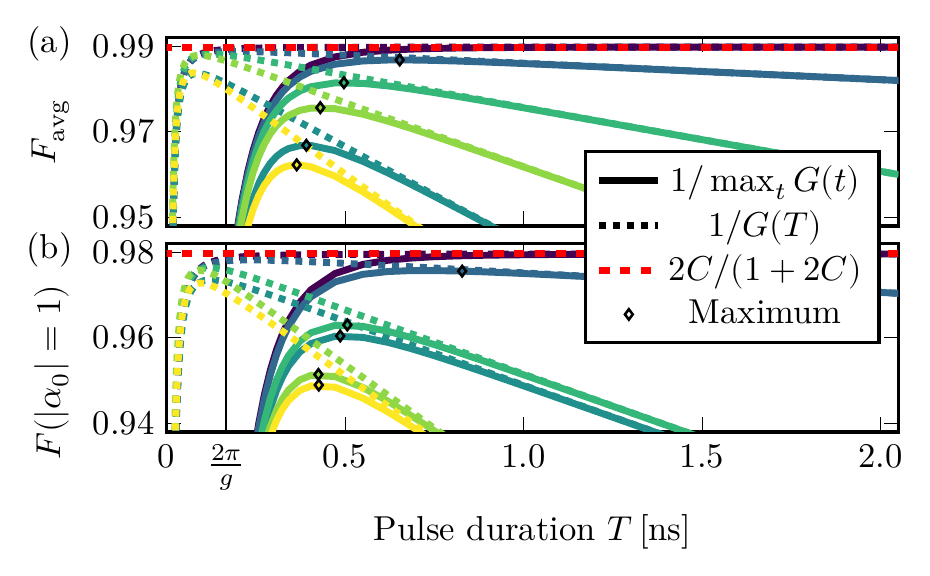} 
\caption{\label{fig:bound}
  Bounds for the average (a) and worst case (b) fidelity as a function of the pulse duration for a \(\sin^2\) pulse, see Eqs.~\eqref{eq:Fid}~and~\eqref{eq:Favg}.
  We compare different bounds,
  derived from different expressions for $E^2$ (see legend),
  the pulse form dependent fidelities based
  on the exact maximum Eq.~\eqref{eq:Emax}
  (solid lines),
  the simplified analytic bound
  \(E^2 = 1/\G{T}\) (dotted lines),
  and the approximation
  for slowly varying pulses and perfect emitters
  $E^2 = 2C / (1+2C)$ 
  [see Eq.~\eqref{eq:slowbound}] (dashed line).
  The maxima of the fidelity based on the exact maximum are marked by diamonds.
  The pulse form dependent bounds are shown for different decoherences of the matter qubit \((\Gamma_1, \Gamma_2)/\tilde{\gamma}\), from dark (blue) to light (yellow) the values are $(0,0)$, $(0.01,0.005)$, $(0,0.1)$, $(0.1,0)$, $(0.2,0)$, and $(0.1,0.1)$.
  We use cavity QED parameters that can describe silicon-vacancy defects in diamond \((g,\kappa,\gamma) = 2\pi\times(6,30,0.1)\,\)GHz inside a perfect one-sided cavity $\tilde{\kappa} = 0$.
}
\end{figure}

A complex square-root in Eq.~\eqref{eq:alpha} would contradict our ansatz such that we find the bound
\begin{align}
  \label{eq:Emax}
  E \le \EM = \frac{1}{\sqrt{\max_{t \ge 0} \G{t}}} .
\end{align}
Because as soon as the square-root in Eq.~\eqref{eq:alpha} tends to zero 
the Rabi amplitude $\Omega(t)$~\eqref{eq:Omega} diverges
(unless $\eta(t), \zeta(t), \dot{\zeta}(t)$ vanish at the same time)
the physical bound can be formulated even stronger such that the inequality becomes strict ($\le \to <$) ensuring $|\alpha(t)| > 0$.
%
The form of $d(\tau)$ [Eq.~\eqref{eq:dG}] and Eq.~\eqref{eq:Emax} imply that a varying phase in the rotating frame,
$\dot{\theta}(t) \ne 0$,
is detrimental, i.e., reduces $\EM$, 
if the phase of $\ket{1}$ can be controlled better than the ES (\(\Gamma_1 < \tilde{\gamma}\)).
Using the method of partial integration,
\(2 \int_0^t e^{(\Gamma_1 - \Gamma_2) t} \left[ \dot{\theta} \ddot{\theta} f^2 + \dot{\theta}^2 f \dot{f} \right]^2 dt = e^{(\Gamma_1 - \Gamma_2) t} f^2(t) \dot{\theta}^2(t) - (\Gamma_1 - \Gamma_2) \int_0^t \dot{\theta}^2 f^2 e^{(\Gamma_1 - \Gamma_2) t} dt\),
when integrating Eq.~\eqref{eq:dG} proves that a varying phase in the rotating frame reduces $\EM$, as long as \(\Gamma_1 < \tilde{\gamma}\).
The bound on the parameter \(E\) also limits the fidelity \eqref{eq:fidelity},
\begin{align}
  F 
  \label{eq:Fid}
  \le & e^{- \Gamma_2 T} \abs{ 1 - (1 - \EM) \abs{\alpha_0}^2 }^{2} .
\end{align}
The fraction of the amplitude transferred from \(\ket{1}\) to \(\ket{1}_v\)
corresponds to the worst case fidelity
$F(|\alpha_0| = 1) = \left| {\lambda(T)}/{\alpha_0} \right|^2 = E^{2} e^{-\Gamma_2 T} \le \EM^2 e^{-\Gamma_2 T}$
and does not depend on the initial condition.
This result is a generalization of the upper bound for the maximum efficiency in Ref.~\cite{vasilev10}
and we emphasize that by incorporating the imperfections of the emitter the fidelity has a maximum at a finite time (see Fig.~\ref{fig:bound}) and is therefore suited to optimize the pulse duration.
Considering that any initial matter-qubit state should be transferred with a good fidelity
we average the fidelity~\eqref{eq:Fid} over the Bloch sphere
\begin{align}
  F_{\mathrm{avg}} 
  =\, & \frac{E^2 + E + 1}{3 e^{\Gamma_2 T}}
  \label{eq:Favg}
    \le  \frac{\EM^2 + \EM + 1}{3 e^{\Gamma_2 T}} .
\end{align}

To compare our result to 
previous bounds for perfect emitters ($\Gamma_1,\Gamma_2=0$) and slowly evolving pulses 
we first
use \(\max_t \G{t} \ge \G{T}\) and 
\(\G{T} > 0\) 
for good quantum emitters where
\(\Gamma_1, \Gamma_2 < \kappa, \tilde{\gamma}\),
employing \(\G{T}\) as a ``simpler'' upper bound for Eqs.~\eqref{eq:Emax}--\eqref{eq:Favg}.
\begin{figure}
\centering
\includegraphics[width=\linewidth]{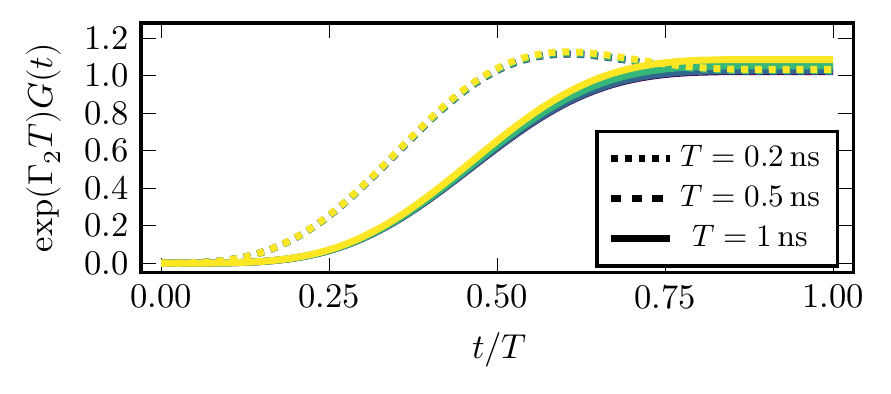} 
\caption[Integrated depletion rate as a function of integration time.]{\label{fig:Gt}
  Integrated depletion rate \(\int_0^{t} d(\tau) d\tau\) [see Eq.~\eqref{eq:dG}] for a \(\sin^2\) puse as a function of integration time \(t\).
  We use this integral to calculate the maximum achievable fidelity for a set of parameters.
  Here, we consider a \(\sin^2\) pulse which corresponds to the ansatz in Eq.~\eqref{eq:fAns} with \(L=1\).
  We use the colors and parameters of Fig.~\ref{fig:bound}.
}
\end{figure}
In Fig.~\ref{fig:Gt} we show $\exp(\Gamma_2 T) \G{T}$ for $\sin^2$ pulses of different duration and for different decoherence rates of the emitter.
The figure shows that for long pulse durations the difference between \(\G{T}\) and \(\max_t \G{t}\) becomes smaller.
It is also readily visible that for short pulses the maximum deviates from \(\G{T}\).
We can further approximate this
for a slowly evolving pulse \(\dot{v} \ll \kappa, g\) and perfect emitter 
to
\begin{align}
  \label{eq:slowbound}
E^2 \lesssim \frac{\kappa}{\kappa + \tilde{\kappa}} \frac{2C}{(1+2C)},
\end{align}
with the (generalized) cooperativity \(C = {2 g^2}/[{\tilde{\gamma} (\kappa + \tilde{\kappa})}]\).
This is in agreement with the photon escape (or  retrieval) efficiency found in 
\cite{gorshkov07,dilley12,muecke13,morin19}.
However, the photon envelope dependent bounds for decohering emitters are more {accurate} and can quantify the optimal duration (see Fig.~\ref{fig:bound}).

To gain a better understanding of our results,
we consider a pulse shape of the form
\begin{align}
  \label{eq:fAns}
  v(t) 
  = \sum_{n=1}^{L} v_n \left[ 1 - \cos\left( \frac{2 \pi n}{T} t \right) \right]
\end{align}
for $0 < t < T$, else $v(t)=0$.
Pulses of this form are real $v(t) \in \mathbb{R}$, symmetric, fulfill \(v(0)=v(T)=\dot{v}(0)=\dot{v}(T)=0\) \cite{martinis14},
and contain \(L\) independent parameters \(T, v_{n}/v_1\) with \(n=2,\dots,L\). 
For \(L=1\) this ansatz is a \(\sin^2\) pulse with variable pulse duration.
Using the basis of Eq.~\eqref{eq:fAns} normalization yields $v_1 = \sqrt{6/9T}$ (\(L=1\))
making it apparent that the only free parameter of the \(\sin^2\) pulse is the duration.

We show the maximum worst-case fidelity \(F(|\alpha_0|=0)\) and average fidelity for \(L=1\) as a function of the pulse duration \(T\) in Fig.~\ref{fig:bound},
confirming that the various approximate bounds fail to capture a useful bound
for the maximum achievable fidelity for all possible timescales and fail to quantify the optimal duration
in contrast to the bound provided by Eq.~\eqref{eq:Emax} that does not rely on an adiabatic approximation and holds for any detuning.

\begin{figure}
\centering
\includegraphics[width=\linewidth]{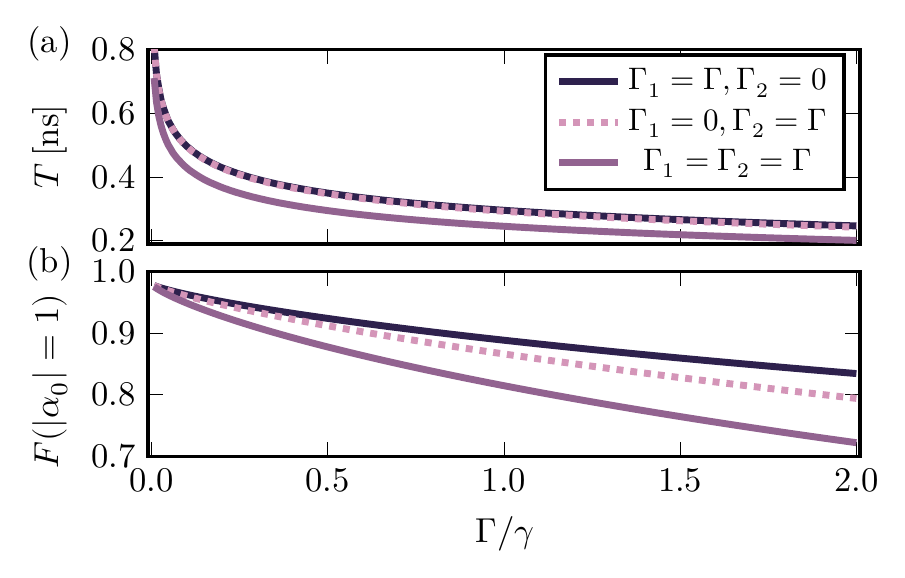} 
\caption[Optimal pulse duration and worst case fidelity of a \(\sin^2\) pulse as functions of the matter qubit decoherence rates $\Gamma_1, \Gamma_2$.]{\label{fig:duration}
  Optimal pulse duration \(T\) for \(L=1\) and worst case fidelity $F(|\alpha_0|=1) = |\lambda(T)/\alpha_0|^2$ as functions of the matter qubit decoherence rates $\Gamma_1$ and $\Gamma_2$ (see legend).
  The colors correspond to different decoherence processes, see legend.
  We use the parameters of Fig.~\ref{fig:bound}. 
}
\end{figure}
The optimal duration balances the finite cavity coupling strength $g$ and cavity decay rate $\kappa$ with the decoherence of the emitter.
We more thoroughly investigate the optimal duration for a \(\sin^2\) pulse
as a function of the qubit decoherence rates \(\Gamma_1\) and \(\Gamma_2\) in Fig.~\ref{fig:duration}.
For this figure we determine the optimal duration numerically in two steps.
First, we find the optimal duration on a grid with 200 time-points between $\max(1/g, 1/\kappa)$ and $\min(\Gamma_1, \Gamma_2)$ and then repeat the optimization with 200 additional points between the two points around the optimum of the first run.
This approach ensures that even for small decoherence rates the grid is sufficiently small to avoid artifacts of the numerical grid.
The decrease of the worst case fidelity as well as the optimal duration as a function of the decoherence rates of the matter system becomes apparent in Fig.~\ref{fig:duration}.
However, the  decrease of the duration flattens for higher rates as the duration is also limited from below due to the finite coupling between the matter-qubit and the cavity and out-coupling rate of the cavity.
Furthermore, we see that qualitatively the rates \(\Gamma_1, \Gamma_2\) behave similarly,
but depending on the protocol of interest it might be preferential to swap the roles of \(\ket{0}\) and \(\ket{1}\),
i.e., couple the shorter or longer lived state to the excited state via the cavity.


\subsection{Fidelity Optimization by Pulse Shaping}\label{sec:stimopt}
\begin{figure*}[t]
\centering
\includegraphics[width=\linewidth]{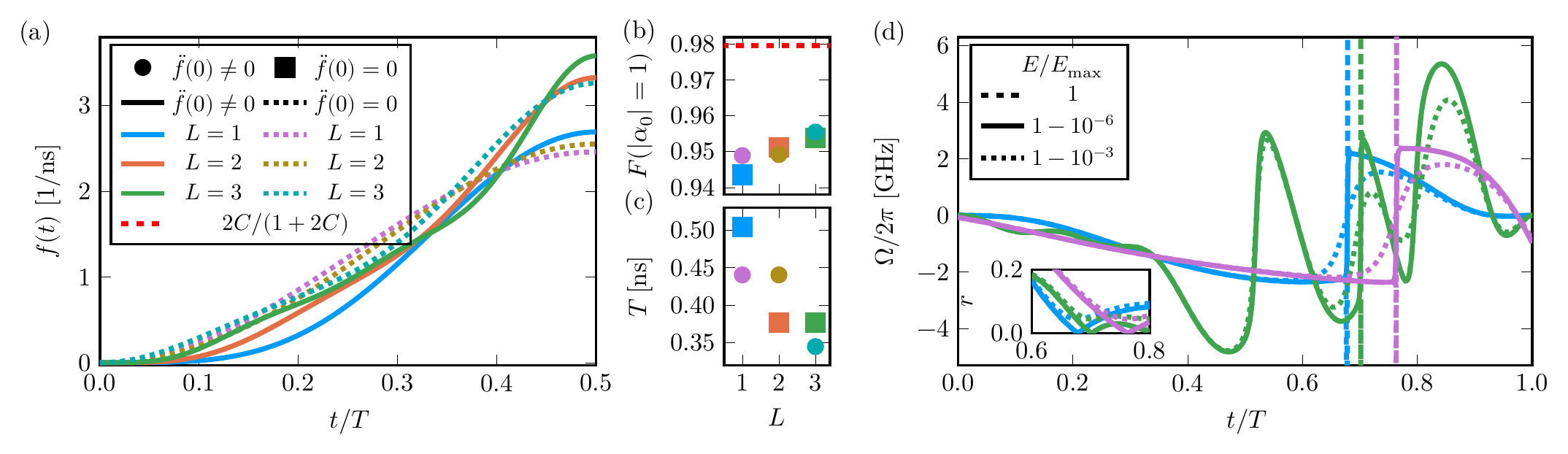} 
\caption[Results of optimizing the pulse shape for a maximum worst-case fidelity.]{\label{fig:opt}
  Results of optimizing the pulse shape for a maximum worst-case fidelity.
  Panel (a) shows half of the optimal symmetric pulse shapes $f(t)$ for different numbers of free parameters \(L\) and restrictions on the pulse (see legend).
  The corresponding worst-case fidelity \(F(|\alpha_0|=1)\) and optimal pulse duration $T$ are depicted in panels (b) and (c), here the dashed line corresponds to the perfect emitter bound [see Eq.~\eqref{eq:slowbound}].
  (d) Driving Rabi frequency \(\Omega(t)\) as a function of time for different pulses [colors see (a)] and \(E/\EM\) (see legend).
  Reducing \(E\) leads to a larger minimal \(\abs{\alpha(t)} = e^{-\Gamma_2 t}r(t)\) (see inset) and smoothes out discontinuities in \(\Omega(t)\).
  Parameters are those of Fig.~\ref{fig:bound}, \(\Gamma_1, \Gamma_2 = 0.1 \gamma\), and \(\Delta = 0\).
  See App.~\ref{app:pulseshapes} for the optimization results.
}
\end{figure*}

More generally,
we can take an ansatz for the envelope and optimize its independent parameters with regard to
any quantity limited by Eq.~\eqref{eq:Emax}.
Here, we maximize
the worst-case fidelity \(F(|\alpha_0|=1)\) [see Eq.~\eqref{eq:Fid}] to find
  \(v_{\mathrm{opt}} = 
  \mathrm{argmax}_v 
  \left[1/\max_{t} e^{\Gamma_2 T} \G{t}\right]\).
The paradigm shift from searching for the optimal $\Omega(t)$
to searching the optimal pulse shape $v(t)$ (including the duration)
saves one from repeatedly solving the dynamics and instead allows the use of the closed-form function $d(t)$ [Eq.~\eqref{eq:dG}], see App.~\ref{app:Ganalytic} for the analytic evaluation of \(\G{t}\) for the ansatz in Eq.~\eqref{eq:fAns}.
Parametrizing $v(t)$ by $\vec{v}$ the optimization is over a finite parameter space instead of a function space,
and the optimization assumes the well-known form of a (continuous) minimax problem \cite{rustem98,rustem09}
for finding
the optimal 
\(\vec{v}\).
In Fig.~\ref{fig:opt} we show results for the ansatz in Eq.~\eqref{eq:fAns} and a grid optimization to find the \(L\)
optimal parameters \(T,v_n/v_1\) (\(n=2,\dots,L\)).
It turns out that increasing $L$ allows for higher fidelities and shorter pulses.
We also display the photon shape and Rabi frequency
to both of which we have immediate access after the optimization.
We remark that the optimization results can be confirmed using a single simulation of the Lindblad master equation with the optimal photon shape and Rabi frequency.

Constraints can be accounted for by different $v(t)$, e.g., by fixing some parameters;
in particular, if the rate at which the drive can be modulated is larger than the optimal \(T\)
we can fix it to the minimal achievable \(T\).
In Fig.~\ref{fig:opt} we also show the optimization results ensuring a continuous drive activation by imposing \(\ddot{v}(0) = 0\) [see Eqs.~\eqref{eq:zeta}~and~\eqref{eq:Omega}].
To this end we only use odd \(n\) as free parameters
and set \(v_{2n} = - \frac{n^2}{(n+1)^2} v_{2n-1}\).
Additionally, we show in Fig.~\ref{fig:opt}(d) that
targeting $\EM$ leads to a pole in the driving Rabi frequency $\Omega(t)$~\eqref{eq:Omega} which can be avoided by smoothing $\Omega(t)$ by reducing the target efficiency \(E < \EM\).

\section{Conclusion}\label{sec:stimconc}

In conclusion,
we derived a distinct bound for the fidelity for cavity-assisted stimulated Raman emission, taking into account not only the cavity quantum electrodynamic quantities, but also the temporal pulse shape and additional decoherence processes of the three-level system 3LS.
Due to the cascaded nature of the equations, we see great potential in applying the bounds to non-trivial waveguides.
Furthermore, we showed how this new bound can be cast into an optimization problem for the pulse shape for an efficient emission process {providing a paradigm shift from optimizing the drive to optimizing the temporal mode (and thereby also fixing the drive).
We show that this optimization is of a closed form expression in contrast to the na\"ive optimization where the dynamics need to be solved (numericaly) repeatedly}.
Combined with the encoding and entanglement protocols we propose,
this is a promising 
ingredient for quantum technology.
The method of including the main emission process into the coherent dynamics by combining the novel input-output approach by \cite{kiilerich19,kiilerich20} with the temporal mode matching can potentially be applied to many problems.
While we grid-optimized a symmetric pulse, 
non-symmetric pulses compatible with the initial conditions,
i.e., continuously vanishing at $t=0$,
and different optimization algorithms
can be investigated analogously.
A natural next step would be to study a photon mediated matter-to-matter transfer within this framework.


\begin{acknowledgments}
  We acknowledge funding from the European Union’s Horizon 2020 research and innovation programme under Grant Agreement No.~862721 (QuanTELCO),
  as well as from the German Federal Ministry of Education and Research (BMBF) under the Grant Agreement No.~13N16212 (SPINNING).
\end{acknowledgments}

\appendix
\section{Rotating Frame and Lab Frame}\label{app:LF}

Since 
the main text introduces the Hamiltonian in the rotating frame, in this section we link the rotating frame to the lab frame.
The Hamiltonian in the lab frame is
\begin{align}
  \tilde{H}_S / \hbar
  =\, & (\Delta + \omega_c) \proj{e} - \delta \proj{1} + \omega_c c^{\dag} c \notag \\
  \label{app:eq:Hlab}
      & + \left(\tilde{\Omega}^{*}(t) \op{1}{e} + g c^{\dag} \op{0}{e} + \mathrm{H.c.} \right) ,
\end{align}
with the excited state energy \(\Delta + \omega_c\) (in units of frequency),
the qubit level splitting \(\delta\),
the cavity frequency \(\omega_c\).
We then apply the transformation to the rotating frame
\begin{align}
  \label{eq:U}
  U = \exp \left[ - i ( - \delta \proj{1} + \omega_c \proj{e} + \omega_c c^{\dag} c) t \right] ,
\end{align}
leading to
\begin{align}
  H_s & / \hbar = U^{\dag} \tilde{H}_S / \hbar U - i U^{\dag} \dot{U} \notag \\
  \label{app:eq:HS}
    & = \Delta \proj{e} + \left(\Omega^{*}(t) \op{1}{e} + g c^{\dag} \op{0}{e} + \mathrm{H.c.} \right) ,
\end{align}
corresponding to the Hamiltonian
of the main text.
The detuning is \(\Delta\) and the drive in the lab frame is linked to the rotating frame by
\(\tilde{\Omega}(t) = \Omega(t) \exp[- i (\delta + \omega_c) t]\).
Additionally, we see how the splitting between the qubit states is absorbed in the time-dependent drive.

Using the same transformation we can also link the time-dependence of the coupling to the virtual cavity \(\tilde{g}_v(t) = g_v(t) e^{i \omega_c t}\)
between the lab and rotating frames.
This also links the photon envelopes \(\tilde{v}(t) = e^{- i \omega_c t} v(t)\)
and leads to the correct transformation of the dissipator $L_0$.
Because the remaining dissipators only gain a global phase in the rotating frame and always occur in pairs (with their adjoint) in the master equation,
we can treat the remaining dissipators as unchanged in the rotating frame.

\section{Decoherence Rates}\label{app:decrate}
In this section we show how to derive the combined decoherence rates arising in the non-Hermitian Hamiltonian from established dissipators from the Lindblad master equation.
The Lindblad master equation takes the form
\begin{align}
\label{app:eq:me}
\frac{d\rho}{dt} = - \frac{i}{\hbar} [H,\rho]+ \sum_{i}\left( L_i \rho L_i^\dagger-\frac{1}{2}\left\{L_i^\dagger L_i, \rho\right\}\right),
\end{align}
also found in the employed open quantum systems approach to input-output theory \cite{kiilerich19,kiilerich20}.
In addition to the main dissipator \(L_0\) introduced in the main text we consider the following dissipators:
\begin{itemize}
  \item Decays from the excited state (ES) \(\ket{e}\) to the ground states (GSs) \(\ket{1}, \ket{0}\): \\
    \(L_1 = \sqrt{\gamma} \cos(\xi) \op{1}{e} \), \(L_2 = \sqrt{\gamma} \sin(\xi) \op{0}{e} \)
  with the branching angle $\xi$
  \item The important uncorrelated dephasing terms~\cite{li12}, here the dephasing of \(\ket{1}\) and \(\ket{e}\) using the phase of \(\ket{0}\) as reference:
    \(L_{3} = \sqrt{\Gamma_{\mathrm{ph}}^1} \proj{1}\) and \(L_{4} = \sqrt{\Gamma_{\mathrm{ph}}^e} \proj{e}\)
  \item Incoherent transitions between the GSs:
\(L_{5} = \sqrt{\Gamma_{\mathrm{0\to1}}} \op{1}{0}\), \(L_{6} = \sqrt{\Gamma_{\mathrm{1\to0}}} \op{0}{1}\)
  \item Unwanted cavity losses (e.g., losses through the wrong mirror), given by \(L_{7} = \sqrt{\tilde{\kappa}} c\)
\end{itemize}

The effect of these dissipators on the non-Hermitian Schr\"odinger equation
 \begin{align}
  \label{app:eq:NHSEQ}
  i \hbar \frac{\partial}{\partial t} \ket{\Psi} = H_{\mathrm{NH}} \ket{\Psi} = \left( H - \hbar \frac{i}{2} \sum_i L_i^{\dag} L_i \right) \ket{\Psi} ,
\end{align}
only depends on \(L_i^{\dag} L_i\).
This implies we can combine
\(\sum_{i = 1,2,4} L_i^{\dag} L_i = \tilde{\gamma} \proj{e}\) with
\(\tilde{\gamma} = \gamma + \Gamma_{\mathrm{ph}}^e\)
and
\(\sum_{i=3,6} L_i^{\dag} L_i = \Gamma_1 \proj{1}\) with
\(\Gamma_1 = \Gamma_{1 \to 0} + \Gamma_{\mathrm{ph}}^1\).
In the main text we refer to \(\Gamma_{0 \to 1} = \Gamma_2\) for a consistent and simpler notation.
Lastly the cavity losses lead to $L_7^{\dag} L_7 = \tilde{\kappa} c^{\dag} c$.

\section{Solving the Non-Hermitian Dynamics}\label{app:solvingHNH}

Combining the temporal mode matching and the non-Hermitian Schr\"odinger equation the (decaying) dynamics of the coherent state transfer are
\begin{align}
  \label{eq:dalpha}
\dot{\alpha}(t) =\, & - \frac{\Gamma_{1}}{2} \alpha(t) + \Omega^{*}(t) \zeta(t), \quad
\dot{\beta}(t) =  - \frac{\Gamma_{2}}{2} \beta(t), \\
  \label{eq:dzeta}
\dot{\zeta}(t) =\, & \left(- i \Delta - \frac{\tilde{\gamma}}{2} \right) \zeta(t) - g \eta(t) - \Omega(t) \alpha(t), \\
  \label{eq:deta}
\dot{\eta}(t) =\, & - \frac{\Gamma_{2} + \kappa + \tilde{\kappa}}{2} \eta(t) + g \zeta(t),  \\
  \label{eq:dlambda}
\dot{\lambda}(t) =\, & - \frac{\Gamma_{2}}{2} \lambda(t) + \frac{\kappa {|\eta(t)|}^{2}}{2 \lambda^{*}(t)},
\end{align}
where we combined important uncorrelated dephasing terms~\cite{li12} and relevant decays from the states $\ket{1}$, $\ket{0}$, and $\ket{e}$ in the rates \(\Gamma_1\), \(\Gamma_2\), and \(\tilde{\gamma}\), respectively.

To solve these equations we first formally integrate the pulse amplitude
\begin{align}
  \label{eq:lambdaFormal}
  \lambda(t) = \sqrt{\kappa} e^{i \varphi} \sqrt{\int_{0}^{t} e^{\Gamma_{2} (\tau-t)} {|\eta(\tau)|}^2\, d\tau} ,
\end{align}
where we allow for an arbitrary initial phase \(\varphi\).
Inserting this solution into the temporal mode matching
\(g_v(t) = - {\sqrt{\kappa} {\eta^{*}{(t)}}}/{\lambda^{*}{(t)}}\)
and comparing to the definition of the time dependent coupling to the virtual cavity
\(g_v(t) = -{v^*(t)}/{\sqrt{\int_0^t dt'\, |v(t')|^2}}\)
relates \(\eta(t)\) to the pulse shape
\begin{align}
  \label{eq:eta-app}
  \eta(t) = E \abs{\alpha_0} e^{i \varphi} e^{-\Gamma_2 t / 2} v(t) / \sqrt{\kappa} ,
\end{align}
with \(\alpha_0 = \alpha(0)\).
We term the positive proportionality constant \(E\) the {\em matter-photon conversion efficiency} and demonstrate in the main text, that it is directly related to the fidelity of the process.
Eq.~\eqref{eq:eta-app} shows that the pulse shape is directly linked to the dynamics of the ES and
combined with the temporal mode matching we see that a static phase (such as \(\varphi\)) can be either encoded in \(\lambda(t)\) and or \(v(t)\).
Reinserting $\eta(t)$ into the formal integral describing the photon probability amplitude yields
\begin{align}
  \label{eq:lambda-app}
   \lambda(t) = E \abs{\alpha_0} e^{i \varphi - \Gamma_2 t / 2} \sqrt{ \int_{0}^{t} \abs{v(\tau)}^2 \, d\tau } .
\end{align}

Inspired by results on the optimal control to emit a photon of a certain shape \cite{vasilev10},
we solve the remaining equations \emph{in reverse} by imposing a fixed photon shape [and thereby \(\eta(t),\lambda(t)\)].
This approach leads us to
\begin{align}
  \label{eq:zeta-app}
  \zeta(t) = \frac{\Gamma_{2} + \kappa + \tilde{\kappa}}{2 g} \eta(t) + \frac{1}{g} \dot{\eta}(t) ,
\end{align}
and the optimal drive
\begin{align}
  \label{eq:Omega-app}
  \Omega(t) = - \frac{( \tilde{\gamma}/2  + i \Delta ) \zeta(t) + g \eta(t) + \dot{\zeta}(t)}{\alpha(t)} .
\end{align}
Because we can ensure that \(| \alpha(t) | > 0\) for \(0 \le t < T\) if \(\alpha_0 \ne 0\)
and the form of \(\Omega(t)\) is irrelevant for \(\alpha_0 = 0\) as no emission takes place,
this provides a solution of Eq.~\eqref{eq:dzeta} for $\Omega(t)$.
Additionally, we now choose \(\varphi\) such that \(\Omega(t)\) becomes independent of the initial condition, i.e.\ \(\abs{\alpha_0} e^{i \varphi} = \alpha_0\) implying that the phase of \(\alpha_0\) becomes the phase of \(\lambda(t)\).
The complex phase of \(\Omega(t)\) is then only determined by the system parameters and the pulse shape \(v(t)\), not the initial condition of the matter qubit.

To separate the remaining Eq.~\eqref{eq:dalpha} into two separable (in $t$) real equations we take the ansatz
\(\alpha(t) = \alpha_0 r(t) e^{i \phi(t) - \Gamma_1 t/2}\),
with \(r(t),\phi(t) \in \mathbb{R}\), \(r(0) = 1\), and \(\phi(0)=0\).
The separation yields
\(r \dot{r} = - \frac{1}{2} E^2 d(t)\)
with the depletion rate accoring to Eq.~\eqref{eq:dG} of the main text.
We can integrate both sides of the separated equation for \(\dot{r}\) to calculate the solution used in the main text for \(r^2(t) =  1 - E^2 \G{t} \ge 0\).
Note that \(r(t)\) and \(d(t)\) are independent of the detuning \(\Delta\).

After solving this integral the phase evolution also becomes a straight-forward integral, i.e.
\begin{widetext}
\begin{align}
  \label{eq:dphi}
  \phi(t) =\, & \int_0^t \frac{F^2 e^{(\Gamma_1-\Gamma2) t} }{g^2 r^2(t)} \Bigg\{
             \left[ \left( 1 + \frac{\tilde{\kappa}}{\kappa} \right) \left( \Delta + \dot{\theta}(t) \right) + \frac{\ddot{\theta}(t)}{\kappa} \right] f(t) \dot{f}(t)
    + \frac{\Delta + 2 \dot{\theta}(t)}{\kappa} \dot{f}^2(t)
    - \frac{\dot{\theta}(t)}{\kappa} f(t) \ddot{f}(t)
             \notag \\
    & + \left[ \frac{\kappa}{4} \left( 1 + \frac{\tilde{\kappa}}{\kappa} \right)^2 \left( \Delta + \dot{\theta}(t) \right) + \left( 1 + \frac{\tilde{\kappa}}{\kappa} \right) \frac{\ddot{\theta}(t)}{2}
    + \frac{1}{\kappa} \left( \Delta \dot{\theta}^2(t) - g^2 \dot{\theta}(t) + \dot{\theta}^3(t) \right) \right] f^2(t)
  \Bigg\} dt ,
\end{align}
with the photon envelope phase \(\theta(t) \in \mathbb{R}\) and amplitude \(f(t) \in \mathbb{R}\), i.e., \(v(t) = e^{i \theta(t)} f(t)\).
We can read of from this expression that \(\phi(t)=\phi(0)=0\) for \(\Delta=0\) and \(\dot{\theta}(t)=0\),
implying that the phase \(\phi\) is constant for \(\Delta=0\) and \(\dot{\theta}(t)=0\),
i.e., a resonant cavity and a pulse shape with constant complex argument.


\section{Analytic Solution of the Integrated Depletion Rate}\label{app:Ganalytic}
In this section we show the analytic expression for $G(t) = \G{t}$ for the real pulse shape introduced in the main text given by
\begin{align}
  \label{eq:fAns_app}
  v(t) = f(t) = \sum_{n=1}^{L} v_n \left[ 1 - \cos\left( \frac{2 \pi n}{T} t \right) \right]
   = \sum_{n=1}^{L} v_n f_n(t) .
\end{align}
Because this pulse shape is real we can write
\begin{align}
  G(t) = \int_0^t & e^{t \left(\Gamma_{1} - \Gamma_{2}\right)} \Bigg\{
            \left[ 1 +
 \frac{\tilde{\kappa}}{\kappa} + \frac{\tilde{\gamma} - \Gamma_2}{g^{2}} \frac{\kappa}{4} \left( 1 + \frac{\tilde{\kappa}}{\kappa} \right)^2 \right] f^{2}(t)
 + \left[ \frac{1}{g^{2}} \left( 1 + \frac{\tilde{\kappa}}{\kappa} \right) + \frac{\tilde{\gamma} - \Gamma_2}{\kappa g^{2}} \right] \dot{f}^2(t)
 \notag \\
            & + \left[ \frac{2}{\kappa} + \frac{\kappa}{2 g^{2}} \left( 1 + \frac{\tilde{\kappa}}{\kappa} \right)^2 + \frac{\tilde{\gamma}-\Gamma_2}{g^{2}} \left( 1 + \frac{\tilde{\kappa}}{\kappa} \right) \right] f(t) \dot{f}(t)
  \label{eq:dG_real}
              +  \frac{1}{g^{2}} \left( 1 + \frac{\tilde{\kappa}}{\kappa} \right) f(t) \ddot{f}(t)
              + \frac{2}{\kappa g^{2}} \dot{f}(t) \ddot{f}(t)
              \Bigg\} dt ,
\end{align}
where using the ansatz~\eqref{eq:fAns_app} enables us to integrate term wise.
This means the full integral takes the form $\sum_{n,m} v_n v_m X$ where $X$ is made up from products of the prefactor and the integrals shown below.
For brevity of the notation we use $\Gamma = \Gamma_1 - \Gamma_2$ and $\omega_n = \frac{2 \pi n}{T}$ to write the different terms for the integral
\begin{align}
    \label{eq:int_abs2g}
    \int_0^t e^{\Gamma t} f_n(t) f_m(t) dt =\, &
    \begin{cases}
        \frac{u_{m-n}(t,\Gamma) + u_{m+n}(t,\Gamma)}{2}
        + \frac{e^{\Gamma t} - 1}{\Gamma}
        - u_{n}(t,\Gamma)
        - u_{m}(t,\Gamma)
        \text{ for } \Gamma \ne 0 , \\
    \frac{3 t}{2} + \frac{\sin(2\omega_{m} t)}{4\omega_{m}}
    - 2 \frac{\sin(\omega_{m} t)}{\omega_{m}}
        \text{ for } \Gamma = 0 \text{ and } n = m , \\
        \frac{u_{m-n}(t,\Gamma=0) + u_{m+n}(t,\Gamma=0)}{2}
        - u_{n}(t,\Gamma=0)
        - u_{m}(t,\Gamma=0)
        + t
        \text{ else} ,
    \end{cases}
    \\
  \label{eq:int_gdg}
    \int_0^t e^{\Gamma t} f_n(t) \dot{f}_{m}(t) dt =\, &
    \omega_m
    \begin{cases}
    \frac{1 - \cos(\omega_m t)}{\omega_m}
- \frac{1 - \cos(\omega_{2m} t)}{2 \omega_{2m}}
    \text{ for } \Gamma = 0 \text{ and } n = m , \\
        u_m(t) + \frac{u_{m+n}(t,\Gamma) + u_{m-n}(t,\Gamma)}{2}
    \text{ else} ,
    \end{cases}
    \\
    \label{eq:int_dgdg}
    \int_0^t e^{\Gamma t} \dot{f}_{n}(t) \dot{f}_{m}(t) dt =\, &
    \frac{1}{2} \omega_n \omega_m
    \begin{cases}
     t - \frac{\sin(2 \omega_m t)}{2 \omega_m}
     \text{ for } \Gamma = 0 \text{ and } n=m , \\
    h_{m+n}(t,\Gamma) - h_{m+n}(t,\Gamma)
    \text{ else} ,
    \end{cases}
    \\
    \label{eq:int_gddg}
    \int_0^t e^{\Gamma t} f_n(t) \ddot{f}_{m}(t) dt =\, &
    \frac{1}{2}\omega_m^2
    \begin{cases}
    2 h_m(t,\Gamma=0) -
       t + \frac{\sin(2\omega_{m} t)}{2 \omega_{m}}
     \text{ for } \Gamma = 0 \text{ and } n=m , \\
    2 h_{m}(t,\Gamma)
    - h_{m+n}(t,\Gamma)
    - h_{m+n}(t,\Gamma)
    \text{ else} ,
    \end{cases} \\
    \int_0^t e^{\Gamma t} \dot{f}_{n}(t) \ddot{f}_{m}(t) dt =\, &
    \frac{1}{2} \omega_n \omega_m^2
    \begin{cases}
    \frac{1 - \cos(\omega_{2m} t)}{\omega_{2m}}
     \text{ for } \Gamma = 0 \text{ and } n=m , \\
     u_{m+n}(t,\Gamma)
     - u_{m+n}(t,\Gamma)
    \text{ else} ,
    \end{cases}
\end{align}
with
\begin{align}
  h_m(t,\Gamma) = &\, \frac{1}{\Gamma^2 + \omega_{m}^2}\{
    e^{\Gamma t} [ \omega_{m} \sin(\omega_{m} t) + \Gamma \cos(\omega_{m} t) ] - \Gamma
    \} , \\
u_m(t,\Gamma) = &\, \frac{1}{\Gamma^2 + \omega_{m}^2}
        \left\{
        \omega_{m}
        +
        e^{\Gamma t}
        \left[
        \Gamma \sin(\omega_{m} t) - \omega_{m} \cos(\omega_{m} t)
        \right]
        \right\} .
\end{align}
\end{widetext}
\section{Optimized Pulse Shapes}\label{app:pulseshapes}
To optimize the pulse shapes in Fig.~3 we use 500 samples for $T$ between $\max(1/\kappa, 1/g)$ and $\min(1/\Gamma_1, 1/\Gamma_2)$ (here $\approx 0.03\,$ns and $16\,$ns) and for each $\lambda_n/\lambda_1$ ($n = 2, \dots, L$) 200 samples between $-1$ and $1$.
With this we can numerically determine the maxima of these discrete points.
Table~\ref{tab:optimization} shows the optimization results of Fig.~3 of the main text.
\begin{table}[t]
    \centering
    \begin{tabular}{c|c|l|l|l|l|l|l|l|l}
       $L$ & $\ddot{f}(0) = 0$ & $\EM$ & $T$ & $v_1$ & $v_2$ & $v_3$ & $v_4$ & $v_5$ & $v_6$ \\
       \hline
        1 & yes & 0.987 & 0.50 & 1.35 & -0.34 &  &  &  &  \\
 2 & yes & 0.987 & 0.38 & 1.5 & -0.38 & 0.16 & -0.09 &  &  \\
 3 & yes & 0.988 & 0.38 & 1.44 & -0.36 & 0.27 & -0.15 & 0.08 & -0.06 \\
 1 & no & 0.988 & 0.44 & 1.23 &  &  &  &  &  \\
 2 & no & 0.988 & 0.44 & 1.28 & -0.07 &  &  &  &  \\
 3 & no & 0.988 & 0.34 & 1.46 & -0.30 & 0.17 &  &  &  \\
    \end{tabular}
    \caption[Optimized pulse parameters for stimulated Raman emission.]{Optimized pulse parameters. The durations are rounded to two (three for $E_{\mathrm{max}}$) decimal digits. The duration $T$ is given in ns and the amplitudes in units of $1/$ns.}
    \label{tab:optimization}
\end{table}

\bibliography{refs.bib}


\end{document}